\documentclass[journal=jacsat,manuscript=article]{achemso}
\usepackage{chemformula} 
\usepackage[T1]{fontenc} 
\usepackage{xcolor}
\usepackage{hyperref}
\usepackage{color}

\author{G. Sudha Priyanga}
\email{priyanka@unimore.it}
\affiliation{Department of Sciences and Methods for Engineering, University of Modena and Reggio Emilia, 42122 Reggio Emilia, Italy.}
\affiliation{Department of Metallurgical and Materials Engineering, Indian Institute of Technology Madras, Chennai, 600036, India.}
\author{Sushant Kumar Behera}
\email{sushant@lbl.gov}
\affiliation{Materials Science Division, Lawrence Berkeley National Laboratory, CA - 94720, USA.}
\affiliation{Materials Engineering, Indian Institute of Science, Bangalore - 560012, India.}

\title[An \textsf{achemso} demo]
  {Interfacial Dynamics and Catalytic Behavior of Single Ni Atom Site} 


\begin{document}

\begin{abstract}
Single-atom catalysts (SACs) have garnered significant interest due to their ability to reduce metal particles to the atomic scale, enabling finely tunable local environments and enhanced catalytic properties in terms of reactivity and selectivity. Despite this potential, their application has largely been confined to small-molecule transformations as metal-catalyzed reaction. In this study, we present a diverse single-atom nickel (Ni) catalyst established via a nanoporous carbon (NPC) supported practice. This catalyst represents a breakthrough by achieving the bond formation between carbon and nitrogen and interfacial dynamics in the SAC. The present first principle-based density functional simulations establish the reaction dynamics and catalytic behaviour of such SAC. This dynamic nature comprises an exclusive nitrogen intercalated site showing excellent base effects. This base quickly tunes the interfacial atmosphere, enabling dynamic movement of adatoms into the NPC species, significantly changing the reaction path in Ni SACs due to superior steric effects. The research demonstrates that SACs can extend the capabilities of catalytic systems to include a wider range of complex reactions, offering substantial promise for the development of new, efficient synthetic methods for creating value-added molecular products.
\end{abstract}

\maketitle

\textit{Introduction}-~The use of noble metals supported on metal oxides as heterogeneous catalysts is prevalent in industry, with the performance of these catalysts being highly dependent on the size of the metal particles \cite{DONG2024114133,chemcatchem2023,chemrev2018}. This size influences both the efficiency of metal atom usage and the selectivity of the catalyst. To optimize these factors, precise control of particle size through advanced preparation techniques is essential. Single-atom catalysts (SACs), which reduce metal particles to the atomic scale, have garnered significant interest in the field of catalysis, largely due to the pioneering work of Qiao's group \cite{Qiao2011}. While there has been extensive research on noble metals such as Pt, Au, Ir, and Pd in SACs, the exploration of more cost-effective transition metals remains limited. This is despite the concept of isolated active sites within solid catalysts being established earlier \cite{YAO2023107275,LI2022154828}. Transition metals like Ni are particularly important for ${CO}_2$ activation and hydrogenation reactions. Recent studies have highlighted the importance of Ni cluster sizes in determining reaction stability and selectivity \cite{WEI2024872}. SACs provide atomic-level precision, enabling highly tunable local environments and superior catalytic properties, including enhanced reactivity and selectivity. However, despite these benefits, SACs have primarily been used for small-molecule transformations. Extending their application to more complex reactions, such as metal-catalyzed cross-coupling, which is crucial for synthesizing a variety of chemical products, remains a significant challenge \cite{jacs2023}.\\

In this study, we present a groundbreaking development in the field of single-atom catalysts (SACs) by creating a heterogeneous single-atom nickel (Ni) catalyst through a novel supercritical nanoporous carbon (NPC) assisted method. This innovative catalyst enables, for the first time, C-C and C-N bond-forming migratory insertion reactions using SACs, representing a major leap forward in catalyst technology. The successful execution of these complex reactions with SACs expands their potential applications far beyond the conventional small-molecule transformations, paving the way for new possibilities in catalysis.\\

Our quantum mechanical simulations offer crucial insights into the reaction mechanism, uncovering the role of a distinctive nitrogen-rich coordination site. This site demonstrates a surprising base effect, wherein the base temporarily alters the coordination environment, thereby facilitating migratory insertion into an N-C species. This process, which was previously impeded by significant steric hindrance, showcases the ability of SACs to transcend traditional limitations. Furthermore, it highlights the potential of incorporating novel coordination structures, which have been largely unexplored in catalyst design, thus opening new avenues for innovation in the field \cite{acsmaterialsau2022,HERMAWAN2023107410}.\\

The results of this investigation emphasize the immense potential of single-atom catalysts (SACs) in broadening the scope of catalytic systems to encompass a broader array of intricate reactions. Through the successful demonstration of SACs' capability in facilitating C-C and C-N bond formations, this study lays the groundwork for devising novel and efficient synthetic approaches for generating high-value molecular products. These insights not only showcase the versatility of SACs but also point towards their capacity to transform synthetic chemistry by spearheading innovative strides in catalyst design and implementation. 

\textit{Computational Details}-~
Density Functional Theory (DFT) simulations were carried out using the SIESTA code \cite{Soler_2002}. The chosen exchange-correlation (XC) functional encompassed the van der Waals density functional (vdW-DF) and the C09 exchange \cite{PhysRevB.81.161104}, offering enhanced accuracy in line with experimental results, akin to the PBESol XC functional for solids \cite{Yuk2017}. Norm-conserving Troullier-Martins pseudopotentials (PPs) \cite{PhysRevB.43.1993} were applied to account for core electrons of various atomic species. Additionally, valence electrons of specific atomic species like C, N, and Ni were considered in the computational calculations. During geometry optimization and electronic structure computations, a 0.15 eV energy shift was introduced for the polarized double-(DZP) basis set. Structural optimizations were performed with a force convergence limit of less than 0.01 eV/${\AA}$. Notably, the study focused on nanoporous carbon (NPC) system and the interactions between several nitrogen atoms (NPC-N) along with Ni interaction (NPC-Ni) and the reduction of nitrogen in presence of both N and Ni atoms (NPC-N-Ni) system were explored. Various interaction parameters, including density of states (DOS), adsorption energies (AE), and reaction paths were investigated. In the simulations, energy cutoff of 560 eV was utilized for the real space mesh grids in all the systems. Brillouin zone integration employed a $3\times3\times1$ k-point mesh according to the Monckhorst-Pack scheme for both pristine (NPC) and atom-intercalated (NPC-N, NPC-Ni and NPC-N-Ni) systems. The density of states and total energy were calculated using the tetrahedron method. It should be noted that the incorporation of spin-orbit coupling (SOC) was omitted in these calculations to avoid overestimation of energy values in the computed profiles.

\begin{figure*}
\centering
\includegraphics[width=12.0cm,height=12.0cm]{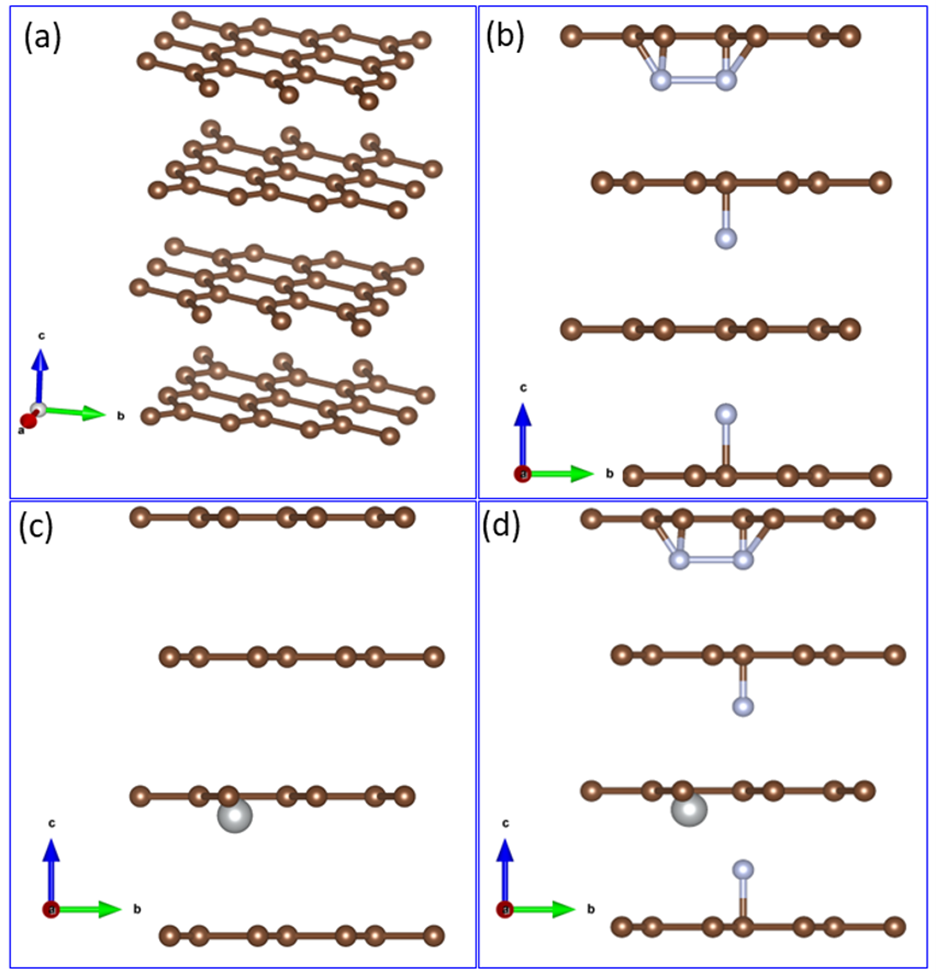}
  \caption{Geometry optimization of systems (a) NPC, and atom-intercalated systems, like (b) NPC-N, (c) NPC-Ni and (d) NPC-N-Ni. Coffee, silver and light blue color for C, Ni and N atoms, respectively. N passivation improves surface activation energy.}
  \label{fig-1}
\end{figure*}

\textit{Results and Discussions}-~The optimized molecular structures of NPC and atom-intercalated systems (NPC-N, NPC-Ni, NPC-N-Ni) are shown in Fig. \ref{fig-1}. Elaborate information regarding the geometric parameters of the simulation cell is provided in Table S1. It's important to note that the bulk density of the molecular system corresponds to the density of the simulation cell. Notably, the density within the simulation cell is adaptable with changes in cell volume, all the while maintaining the structural integrity and independence from external perturbations. In cases of larger simulation cells, residual empty space might persist, which could lead to increased computational overhead. To address this, a complete optimization of the cell volume is implemented, as outlined in Table S2, thereby effectively managing any computational inefficiencies associated with empty space within the simulation cell.

 In this analysis, we have determined the total forces acting on each atom at a fixed radial distance by deriving the total energy with respect to atomic positions within the simulation cell. These computations align with the principles of the Born-Oppenheimer surface, replicating ab initio quantum mechanical simulations. Our findings primarily converge towards the Hellmann-Feynman force equation, focusing on first-order approximations and neglecting higher-order terms (\textit{i.e.}, second and third-degree terms). Here, this presents a breakdown of various energy components comprising ion-electron interactions, incorporating both local and non-local pseudopotential contributions. The non-local aspect of total energy and forces is analyzed on a linear scale. Consequently, this approach enables the effective and efficient handling of larger nanoporous systems through linear scaling Density Functional Theory (DFT) calculations. An intriguing observation from Table S2 is the enhanced dynamism imparted to the NPC system through N passivation, followed closely by the presence of Ni adatoms. This insight sheds light on the mechanisms governing the system's behavior and offers valuable guidance for optimizing its performance in practical applications.

\begin{figure*}
\centering
\includegraphics[width=14.0cm,height=10cm]{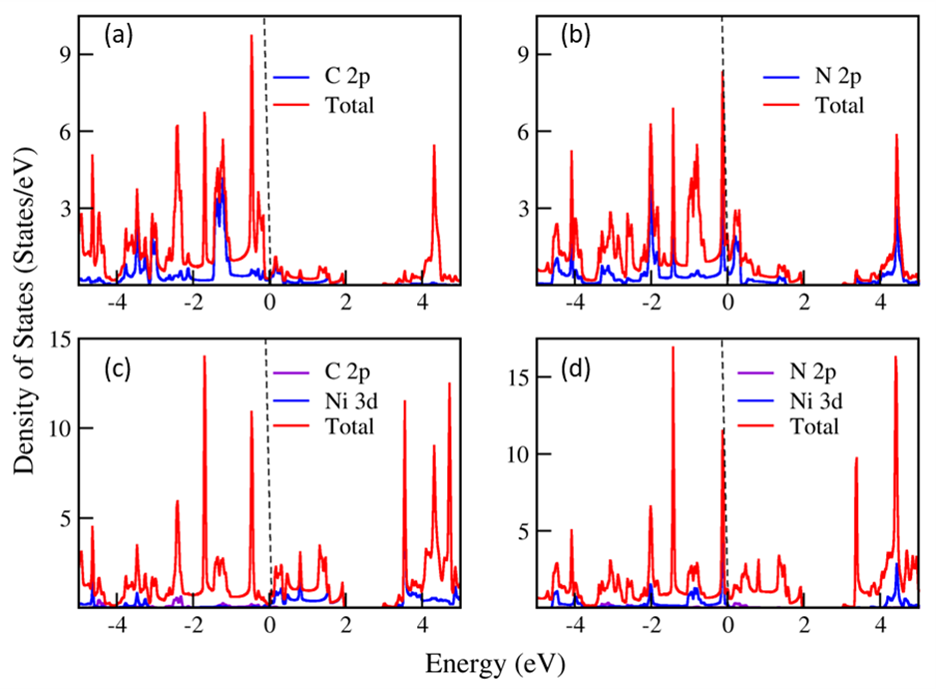}
  \caption{Density of states (DOS) of systems (a) NPC, and atom-intercalated systems, like (b) NPC-N, (c) NPC-Ni and (d) NPC-N-Ni. The vertical dotted black lines indicate the Fermi levels. This dynamic nature of active, delocalized surface electrons is evident on N or Ni atoms are intercalated, or when both N and Ni are simultaneously intercalated.}
  \label{fig-2}
\end{figure*}

\textit{Electronic properties}: Figure \ref{fig-2} provides insight into the electronic density of states (DOS) for the NPC system and all atom-intercalated variations. The DOS patterns confirm heightened activity in proximity to the Fermi level, underscored by the localization of electron states and the discernible influence of atom intercalation on system functionality. Significantly, a considerably broader energy spectrum is observed at the Fermi level when N or Ni atoms are intercalated, or when both N and Ni are simultaneously intercalated. This indicates an amplified adsorption capability and heightened sensitivity towards these specific atomic species. This dynamic nature of active, delocalized surface electrons is distinctly illustrated through the distribution of DOS population density in ion-intercalated systems, resulting in broader distributions with more pronounced peak intensities compared to the pristine NPC system.

\begin{figure*}
\centering
\includegraphics[width=11.0cm,height=8.0cm]{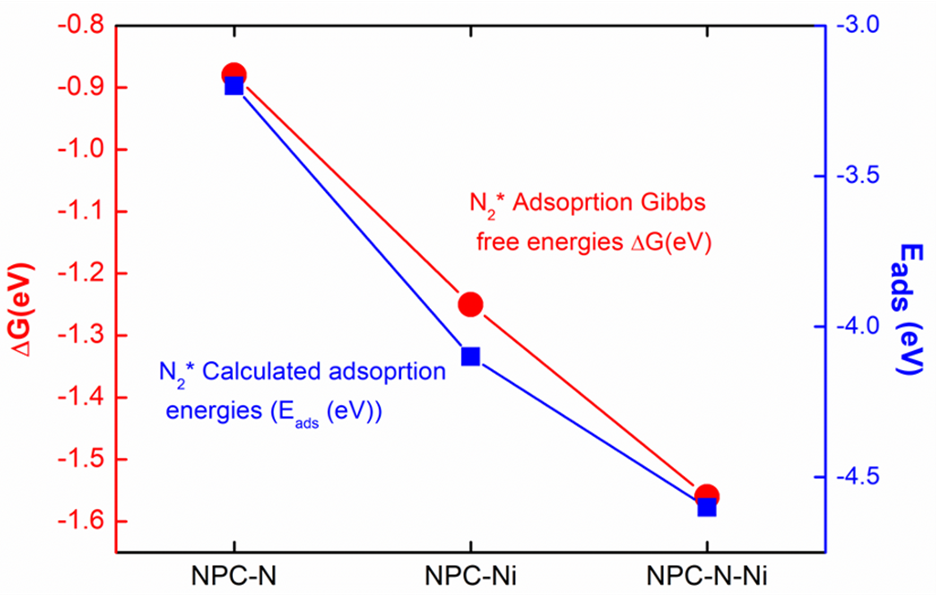}
  \caption{Adsorption Gibbs free energies and calculated adsorption energies of atom-intercalated systems, like NPC-N, NPC-Ni and NPC-N-Ni.}
  \label{fig-3}
\end{figure*}

\textit{Reaction path mechanisms}: According to reports \cite{Pan2022,Wu2020,adma2023Di,acs.chemrev2022}, it is preferable to have a good catalyst with superior adsorption Gibbs free energies ($\Delta$G) and adsorption energies ($E_{ads}$). Figure \ref{fig-3} shows the plot of both Gibbs free energies and calculated adsorption energies for the systems in case of $N_2$* adsorptions. It is observed that the $\Delta$G values of $N_2$ molecule chemisorption on Ni intercalated NPC systems are more negative than those of NPC-N system. Similar trend is also noticed for the $E_{ads}$ values. 

\begin{figure*}
\centering
\includegraphics[width=14.0cm,height=6.0cm]{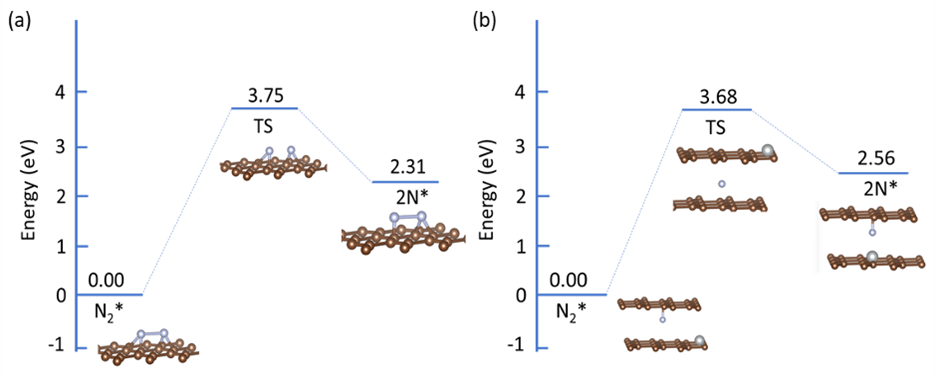}
  \caption{The reaction paths for $N_2$ adsorptions via the $N_2$ dissociation minimum energy path on the NPC system (a) NPC and atom-intercalated system in (b) NPC-Ni.}
  \label{fig-4}
\end{figure*}

In the limelight of activation barrier for the N bonds to form stable 2N via transition state, we further calculate its dissociation barrier energy for the NPC system with N (Fig.~\ref{fig-4}(a)) and Ni (Fig.~\ref{fig-4}(b)) intercalation. The NPC-N system shows a high energy barrier value of 3.75 eV which is still quite difficult for the $N_2$ molecule to dissociate into separate N atoms and later get adsorbed. While the intercalation happens with Ni atom, the bind length got shorter and the barrier reduced to 3.68 eV, making the process smooth. Thus, the metal atom helps in tuning the N bonds exhibiting superior $N_2$ activation capacity.

The Density of States (DOS) patterns indicate an increased level of activity near the Fermi level, highlighting the localization of electron states and the significant impact of atom intercalation on system functionality. Specifically, the intercalation of nitrogen (N), nickel (Ni), or a combination of both results in a notably broader energy spectrum at the Fermi level. This broadening signifies an enhanced adsorption capability and a heightened sensitivity towards these particular atomic species. According to research, an effective catalyst typically exhibits favorable adsorption Gibbs free energies ($\Delta$G) and adsorption energies (${E}_{ads}$). In our study, the plots of Gibbs free energies and calculated adsorption energies reveal the performance of various systems during the adsorption of $N_2$ molecules. For the NPC-N system, the energy barrier for $N_2$ dissociation into separate N atoms is relatively high, at 3.75 eV, which poses a significant challenge for the adsorption process. However, when Ni atoms are intercalated, the binding length is shortened, reducing the energy barrier to 3.68 eV. This reduction in the energy barrier facilitates a smoother adsorption process. Consequently, the presence of Ni atoms aids in tuning the nitrogen bonds, thereby enhancing the system's capacity for $N_2$ activation. This demonstrates that metal atom intercalation, particularly with nickel, plays a crucial role in optimizing the adsorption and activation properties of the system, making it a more effective catalyst for nitrogen-related reactions.

\textit{Conclusions}-~In summary, we introduce a novel single-atom nickel (Ni) catalyst supported by nanoporous carbon (NPC), marking a significant advancement. This catalyst achieves carbon-nitrogen bond formation and showcases dynamic interfacial behavior within the SAC. Utilizing first-principle-based density functional simulations, we uncover the reaction dynamics and catalytic characteristics of this SAC. The dynamic nature of this catalyst is highlighted by an exclusive nitrogen-intercalated site that exhibits exceptional basic effects. This element rapidly adjusts the interfacial environment, facilitating the dynamic movement of adatoms into NPC species. Consequently, this leads to a notable alteration in the reaction pathway within Ni SACs due to superior steric effects. Our study underscores the potential of SACs to expand the repertoire of catalytic systems, allowing for a broader spectrum of complex reactions. This progress holds substantial promise for the advancement of efficient synthetic methods aimed at producing high-value molecular products.

\medskip
\textbf{Acknowledgements} 
 SKB acknowledges DOE, Govt. of USA and UGC, Govt. of India. Parts of the simulations are also performed in Computational facility of SERC, IISc.\\
\textbf{Author Contributions} 
G.S.P and S.K.B. formulated the problem, conducted all the work and analysis. Both the authors wrote the draft.\\
\textbf{Notes} 
The authors declare no competing financial interest.
\bibliography{main}

\end{document}